\documentclass[acmsmall]{acmart}

\usepackage{threeparttable}

\AtBeginDocument{%
  }

\setcopyright{acmlicensed}
\copyrightyear{2025}
\acmYear{2025}
\acmDOI{XXXXXXX.XXXXXXX}

\acmJournal{JACM}
\acmVolume{37}
\acmNumber{4}
\acmArticle{111}
\acmMonth{5}

\newcommand{\interviewquote}[2]{%
    \begin{quote}
        ``\textit{#1}'' (#2)
    \end{quote}%
}

\newcommand{\inlinequote}[1]{%
  ``\textit{#1}''%
}

\begin{document}

\title{Post-Post-API Age: Studying Digital Platforms in Scant Data Access Times}

\author{Kayo Mimizuka}
\email{kayomimizuka@utexas.edu}
\orcid{0000-0003-0966-5386}
\affiliation{%
  \institution{The University of Texas at Austin}
  \city{Austin}
  \state{Texas}
  \country{USA}
}

\author{Megan A Brown}
\email{mgnbrown@umich.edu}
\orcid{0000-0002-1338-8054}
\affiliation{%
  \institution{University of Michigan}
  \city{Ann Arbor}
  \state{Michigan}
  \country{USA}
}

\author{Kai-Cheng Yang}
\email{yang3kc@gmail.com}
\orcid{0000-0003-4627-9273}
\affiliation{%
  \institution{Northeastern University}
  \city{Boston}
  \state{Massachusetts}
  \country{USA}
}

\author{Josephine Lukito}
\email{jlukito@utexas.edu}
\orcid{0000-0002-0771-1070}
\affiliation{%
  \institution{The University of Texas at Austin}
  \city{Austin}
  \state{Texas}
  \country{USA}
}


\begin{abstract}
Over the past decade, data provided by digital platforms has informed substantial research in HCI to understand online human interaction and communication.
Following the closure of major social media APIs that previously provided free access to large-scale data (the ``post-API age''), emerging data access programs required by the European Union's Digital Services Act (DSA) have sparked optimism about increased platform transparency and renewed opportunities for comprehensive research on digital platforms, leading to the ``post-post-API age.''
However, it remains unclear whether platforms provide adequate data access in practice.
To assess how platforms make data available under the DSA, we conducted a comprehensive survey followed by in-depth interviews with 19 researchers to understand their experiences with data access in this new era.
Our findings reveal significant challenges in accessing social media data, with researchers facing multiple barriers including complex API application processes, difficulties obtaining credentials, and limited API usability. These challenges have exacerbated existing institutional, regional, and financial inequities in data access.
Based on these insights, we provide actionable recommendations for platforms, researchers, and policymakers to foster more equitable and effective data access, while encouraging broader dialogue within the CSCW community around interdisciplinary and multi-stakeholder solutions.
\end{abstract}

\ccsdesc[300]{Information systems~Social networks}
\ccsdesc[300]{Social and professional topics~Privacy policies}
\ccsdesc[500]{Social and professional topics~Governmental regulations}
\ccsdesc[500]{Applied computing~Law, social and behavioral sciences}
\keywords{DSA, post-API, social media, data access, survey, interview}


\maketitle

\section{Introduction}

Over the past decade, data provided by digital platforms has informed substantial research in HCI to understand online human interaction and communication.
However, many platform data access programs have shuttered in recent years, leaving researchers with few official means through which to access social media data for research.
In 2023, following Elon Musk's purchase of Twitter (now X), the platform announced its plans to end free access to its Academic API~\citep{clama2023twitter}. Following suit, Meta shut down CrowdTangle in 2024, ending the primary avenue by which researchers accessed and analyzed public information from Facebook and Instagram~\citep{ortutay2024crowdtangle}.
Consequently, many public-interest research projects were interrupted, limiting public understanding of the social media platforms' role in public and civic life.

Around the same time, new regulations in the E.U.'s Digital Services Act (DSA) mandate that Very Large Online Platforms (VLOPs) and Very Large Online Search Engines (VLOSEs) grant researchers access to public data.
These platforms typically comply with the mandates by providing researcher data access programs.
In this paradigm, the companies ultimately decide whether or not a researcher gets access, and each platform may interpret the legal requirements differently.
In light of diminishing public-facing APIs, how platforms grant (or do not grant) access to platform data under the DSA is of utmost importance to the viability of API-based platform scholarship. However, it is unclear whether platforms provide adequate data access in practice.

To assess how platforms make data available under the DSA, we conducted a mixed-method study, combining survey data from 180 responses and in-depth interviews of 19 researchers about their experiences with data access under the DSA.
We find that researchers overall were frustrated with data application processes through DSA-mandated public data access programs.
Our survey shows that many researchers did not apply because they were unaware of the platforms' data access programs, because they found the applications or policies problematic, or because they were not interested in the data offered by the programs.
For those who did apply for data access, the majority had not heard back or had applications rejected at the time of the survey.

The interviews further reveal that researchers are frustrated with the opacity of the data application processes implemented by platforms, with many being denied access without a sufficient reason given by the platform.
Even when researchers did get access to platform data, they found that the APIs were often clunky, unusable, or did not provide the data they needed to conduct their research.
Based on these insights, we provide actionable recommendations for platforms, researchers, and policymakers to foster more equitable and effective data access, while encouraging broader dialogue within the CSCW community around interdisciplinary and multi-stakeholder solutions.

\section{Background}

\subsection{APIs for Research}
Historically, researchers have relied on third parties to collect data, such as firms providing representative panels or organizations facilitating interviews. In social media research, this model has evolved distinctly---instead of traditional third-party research firms, social media companies themselves serve as the intermediaries, primarily through Application Programming Interfaces (APIs).

For these social media companies, user-generated data is a proprietary asset with economic value.
Their data infrastructures are primarily designed to support business objectives rather than academic inquiry~\citep{wu2021platform}.
As a result, the APIs provided by companies are designed to enable external actors to engage with the platform, often in ways that serve the platform's commercial interests.
For example, a business might use the YouTube API to automate video uploads, contributing to the platform's growth.
When companies make APIs publicly available, they typically prioritize commercial applications, with public-interest research emerging as an incidental benefit rather than an intended goal.\footnote{A notable exception was Twitter's Academic API, which, before Elon Musk's acquisition of Twitter in 2022, stood out as one of the first API offerings specifically designed for academic research.}

As social media becomes increasingly central to public and social life, researchers have increasingly depended on platform data access to study the impact of social media on civic, social, and public health.
Previous efforts to study platforms in this manner have yielded important scholarship in HCI and CSCW, including the influence of social media on democratic processes~\citep{praet2021patterns,prochaska2023mobilizing,starbird2019disinformation}, how social media can help amplify or silence marginalized voices~\citep{jackson2020hashtagactivism,wang2023weaving,bhimdiwala2024fighting}, and the role of social platforms in public health crises~\citep{chen2020tracking,karra2023fishing,pater2023social}.
Despite obvious public benefit and harm, platforms often have little incentive to share platform data, especially when findings may reflect poorly on the company~\citep{persily2020social}.

This dynamic results in a condition that Wagner calls ``independence by permission''~\cite{wagner2023independence}.
In an analysis of Meta's research collaborations during the 2020 U.S. election, Wagner illustrates this phenomenon, where access to data depended on corporate approval.
Even researchers who are not directly funded by or partnered with social media platforms remain dependent on platforms' willingness to grant access, effectively placing limits on scholarly independence in social media research.

\subsection{Eras of Data Access}

\begin{figure}[h!]
    \centering
    \includegraphics[width=\linewidth]{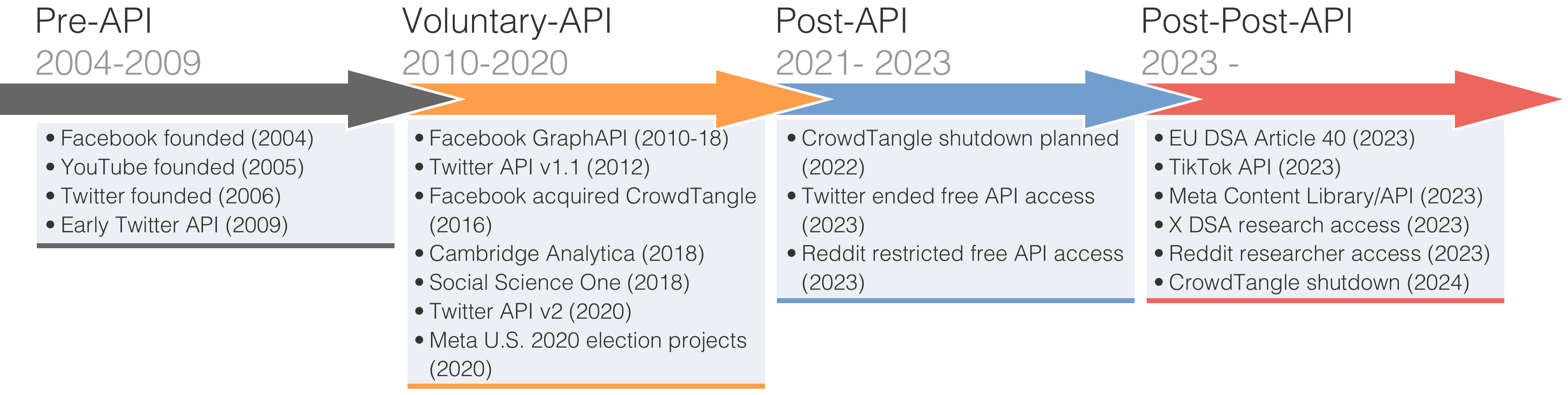}
    \caption{
      Different eras of data access to social media platforms.
      We split the timeline into four eras: the pre-API age, the voluntary-API age, the post-API age, and the post-post-API age.
      We list the key events and developments in each era.
      Note that the time periods are approximate, and different eras and events may overlap.
      }
    \label{fig:timeline}
    \Description[A timeline of the eras of data access to social media platforms.]{A timeline of the eras of data access to social media platforms.}
\end{figure}

Researchers' access to platform data has fluctuated significantly over time.
Here we split the past two decades into four eras and illustrate them in Figure~\ref{fig:timeline} with key events and developments.

\subsubsection{Pre-API and Voluntary-API Eras}

With the emergence of major social media platforms like Facebook (founded in 2004), YouTube (founded in 2005), and Twitter (founded in 2006), both companies and researchers recognized the immense value of digital data as a political and economic resource~\cite{lazer2009computational}.
However, during this early period, which we term the ``pre-API age,'' programmatic approaches to accessing platform data remained limited.

Starting around 2010, researchers experienced a ``golden age'' of data access, as many social media platforms voluntarily developed APIs (the ``voluntary-API age'' in Figure~\ref{fig:timeline}).
One of the most influential examples was Twitter's v1.1 and 2.0 APIs, which were free and open to all users and allowed accessing a wide range of data types with minimal restrictions. The Twitter 2.0 API also had an ``academic track'' that provided increased access for verified academic researchers.
Over the decade, these APIs enabled countless research projects across various academic disciplines~\citep{murtfeldt2024riptwitterapieulogy} and have been used for creating public-service tools and educational resources.
Other researcher access programs from the same era include the Social Science One program~\citep{king2020new}, access through CrowdTangle, and the Meta U.S. 2020 elections Project~\citep{wagner2023independence}.
These programs offered researchers diverse data types and varying levels of access, providing unique insights into different aspects of online platform behavior and user activity.

However, all data access programs have their limitations.
Although APIs offer powerful and flexible access, platforms maintain complete control over the data access process, available data types, and API functionality~\citep{bucher2013objects}.
This means that platforms can modify API specifications or end the API at their discretion, potentially disrupting ongoing research projects and tools.
CrowdTangle also exemplified these constraints---while it provided relatively open access to public posts from Facebook and Instagram, it limited access to content from larger public pages and groups with over twenty-five thousand followers.
The Social Science One initiative fell short of expectations, providing researchers with data of limited utility compared to what was initially promised~\citep{alba2019ahead}.
The project was further compromised when it was discovered that half of the dataset was missing due to production errors~\citep{alba2021facebook}.
At the same time, Facebook maintained control over both CrowdTangle and Social Science One, with the ability to deny data access to researchers whose work might not align with company interests.
Similarly, the Meta U.S. 2020 election projects remained dependent on Facebook's voluntary cooperation to provide data access.
Notably, no subsequent projects with comparable arrangements were implemented or proposed, highlighting the unsustainable nature of this approach.

\subsubsection{The Post-API Age}

The aforementioned limitations collectively undermine the stability and reliability of data access approaches.
A pivotal event in the history of platform data access was the Cambridge Analytica scandal in 2018.
In response to this scandal, Facebook significantly restricted access to its Graph API, dealing a severe blow to the social media research community.
In the influential essay ``Computational Research in the Post-API Age''~\citep{freelon2018computational}, Freelon analyzed the implications of Facebook's decision.
The restrictions on the Graph API effectively eliminated all Terms of Service-compliant methods for systematic collection and analysis of Facebook data at the time.
This shift led Freelon to characterize the resulting era as the ``Post-API'' age for Facebook research.

Another driving factor that led to more restricted data access is the rise of large language models (LLMs).
The development of LLMs has been driven by the so-called ``scaling law,'' which states that the performance of LLMs is a function of the model's size and the amount of training data~\cite{hoffmann2022trainingcomputeoptimallargelanguage}.
This pattern makes user-generated data a valuable resource for training LLMs.
As a result, online platforms have become increasingly cautious about granting data access.

Since 2023, the data access landscape has undergone a series of increasingly restrictive changes.
Twitter initiated this trend by discontinuing its free API access.
While a paid API remains available, its pricing structure has become prohibitively expensive for most researchers, with costs ranging from \$100 per month for 10,000 tweets to \$5,000 per month for one million tweets.
Following Twitter's lead, Reddit implemented similar restrictions on its free API access.
Subsequently, in 2024, Meta announced the discontinuation of its CrowdTangle platform.
These significant changes collectively mark a definitive shift in the ``post-API'' era for social media research.

In response to these challenges, researchers have intensified their exploration of alternative data collection methods beyond platform-provided APIs.
Freelon's essay remains particularly salient, emphasizing that effective social media platform research necessitates diverse data collection strategies, encompassing both collaborative and adversarial approaches.
Many researchers have adopted unsanctioned web scraping as an alternative method~\citep{bruns2021after}.
However, this approach raises significant concerns as it introduces complex ethical and legal challenges that researchers must carefully navigate~\citep{brown2024web}.
As an alternative to scraping, researchers have developed more transparent data collection methods that incorporate explicit participant consent.
For instance, \citet{breuer2023user} demonstrate how researchers can gather Facebook data through voluntary data donations from study participants.
Furthermore, the emergence of open-source tools for facilitating and analyzing user-donated data has established data donation as an increasingly viable research methodology in the post-API era~\citep{araujo2022osd2f}.

\subsection{New Regulatory Avenues in a Post-Post API Age}

While the post-API age has left researchers in limbo, new developments in the European Union signals a new model for data access that is driven by policy regulations and an expectation of transparency for the platforms.
An example is the Digital Services Act (DSA), which introduces a new process where online platforms and search engines provide data access to researchers and the broader public.

Under Article 40.12, platforms and search engines classified as ``very large online platforms'' (VLOPs) or ``very large online search engines'' (VLOSEs) must grant researchers access to publicly available data.
Under Article 40.4, vetted researchers may access private data.
Data access under the DSA is not limited to researchers within the European Union; rather, researchers around the globe can access data through provisions in the DSA provided they meet the same qualifications for data access (namely, that they are affiliated with non-profit institutions and are conducting research that investigates systemic risks to the European Union)~\citep{brown2024web}.

While the DSA represents a positive regulatory development, its implementation ultimately rests with the platforms themselves.
Importantly, under Article 40.12, platforms may still screen researchers for access to public data, introducing many of the same ``independence by permission'' challenges that we highlight in the previous sections.
Moreover, the platforms may interpret and comply with the DSA mandates differently.
Due to these uncertainties, this study aims to understand both the platforms' implementation details and researchers' experiences when applying for data access under the regulation.

\section{Methods}
To study researcher data access, we conducted a mixed-method study involving a survey and subsequent interviews. This process was approved by an Institutional Review Board (protocol number hidden for anonymity).

\subsection{Survey Process}
To conduct the survey, we solicited volunteer participants from seven different professional organizations across the following disciplines: computational social science, communication, political science, internet studies, and human-computer interaction.
We also recruited from the following research communities: the Coalition for Independent Technology Research, the Media and Democracy Data Cooperative, the Knight Research Network, and the Center for Democracy and Technology.
We ran the survey for roughly three months and received 180 responses.

\subsection{Interview Data Collection and Analysis}
After the survey, we conducted a total of 19 semi-structured interviews between October 2024 and February 2025 to gain deeper insight into the researchers' experiences with data access.
We recruited interviewees by reaching out to survey respondents who had indicated their willingness to participate in our interview.
We obtained consent from 17 researchers and recruited two additional interviewees who did not complete the survey but were referred to us by one of the initial interviewees.
It is important to note that some researchers indicated not having API access to a platform during the survey, but gained access between completing the survey and participating in the interview.

Of the 19 participants, 14 (73.7\%) are academic researchers affiliated with universities, 5 (26.3\%) are non-academic researchers affiliated with for-profit companies, public-funded research institutes, or civil-society organizations.
12 (63.2\%) are based in the E.U., 6 (31.6\%) are based in the U.S., and 1 (5.3\%) is based in Latin America.
We have assigned the researchers descriptive codes based on their affiliations: AR for academic research institutions, NR for non-academic institutions, as well as regions they are based in: E.U. for Europe, U.S. for the United States, and L.A. for Latin America.
Some common research fields of the participants include Information Sciences, Communication, Computer Science, Humanities, and Political Science.
Details about the participants are provided in the Appendix, Table~\ref{tab:participant-info}.

The interviews were conducted remotely over Zoom and typically lasted between 45 minutes and an hour.
To refine our interview protocol and ensure that it facilitated relevant discussions, several initial interviews were conducted by two members of the authorship together.
The remaining interviews were conducted with only one author present.
After the interviews, we transcribed them for analysis.
The transcripts of six interviews were generated by Zoom's auto-transcription feature, while the other 13 interviews were transcribed using the Whisper Large v3 Turbo model, released by OpenAI.\footnote{\url{https://github.com/openai/whisper}}
Since the Whisper model often made mistakes and could not accurately label the speakers, an author manually reviewed the transcripts, corrected the errors, and labeled the speakers.

We analyzed the interview transcripts in the following steps.
First, three authors conducted open-coding of several transcripts, reading them line by line to capture relevant themes on a granular level using the qualitative analysis software Atlas.ti.\footnote{\url{https://atlasti.com}}
Next, the authors discussed these initial codes as a team to agree on a final codebook to employ.
During this process, codes with similar meanings were collapsed while the irrelevant codes were removed. We then applied the codebook to all of the transcripts.
Based on the annotation, two authors wrote memos to explain and articulate salient themes and their potential implications. Then, the research team conducted axial coding, discussing relationships between codes and creating larger categories. Finally, we discussed and agreed on the most important themes and featured them in the paper.

\section{Findings}
\subsection{Survey Results}

Our questionnaire focused primarily, though not exclusively, on platforms classified as a VLOP or VLOSE at the time of the survey distribution.
We asked respondents about the data access program of each platform, whether they applied, and for what reasons they chose not to apply.
Note that we included an additional platform, Reddit, owing to its popularity within the U.S. social media ecosystem~\citep{proferes2021studying} even though it was not classified as a VLOP in the DSA.

\begin{table}[h!]
\begin{threeparttable}
\centering
\caption{
  Survey responses about data access across platforms.
  From left to right, we report the number of responses ``I already had access to this data'' (Had access), ``I didn't know I could request data access'' (Unaware), ``I do not believe I am eligible'' (Ineligible), ``I found the application (process) problematic'' (Problematic), and ``I'm not interested in data from this platform'' (Not interested).
  }
\begin{tabular}{lrrrrr}
\hline
\textbf{Platform} & \textbf{Had access} & \textbf{Unaware} & \textbf{Ineligible} & \textbf{Problematic} & \textbf{Not interested} \\
\hline
Alibaba (Ali Express) & 0 & 20 & 2 & 1 & 24 \\
Bing & 0 & 19 & 1 & 0 & 24 \\
Booking.com & 0 & 17 & 1 & 0 & 26 \\
Crowdtangle\tnote{a} & 15 & 6 & 2 & 5 & 2 \\
Google Maps & 0 & 17 & 0 & 1 & 24 \\
Google Play & 0 & 14 & 0 & 0 & 29 \\
Google Records Request & 0 & 5 & 1 & 0 & 6 \\
Google Search & 1 & 22 & 2 & 1 & 19 \\
Google Shopping & 1 & 13 & 2 & 0 & 27 \\
LinkedIn & 0 & 17 & 1 & 1 & 20 \\
Meta Content Library\tnote{a} & 4 & 6 & 4 & 12 & 4 \\
Snap & 0 & 16 & 3 & 1 & 23 \\
TikTok & 1 & 9 & 2 & 10 & 9 \\
X (Twitter) & 4 & 2 & 4 & 15 & 1 \\
YouTube & 7 & 11 & 4 & 5 & 8 \\
\hline
\end{tabular}
\label{tab:dataaccess}
\begin{tablenotes}
    \item[a] Includes Facebook and Instagram.
\end{tablenotes}
\end{threeparttable}
\end{table}

In the survey, applicants could indicate that they did not know that they could request data access, did not believe they were eligible, found the application process problematic, and/or were not interested in data from that particular platform.
We summarize these findings in Table~\ref{tab:dataaccess}.

For many researchers, the primary reason they did not apply for data access from a given platform was because they were not interested in data from that particular platform.
However, researchers also indicated that they did not apply for data access because they were not aware that they could, or that a data access program even existed for a given platform.

There are several notable exceptions to this.
For data access to Meta platforms (Facebook and Instagram), including Crowdtangle and the Meta Content Library, researchers did not apply because they either already had access (through Crowdtangle, which was still available at the time of the survey) or because they were concerned about the application process (for the Meta Content Library).
Similarly, applicants for Twitter's data access program under the DSA were concerned because they found the application process to be time-consuming, unclear, or overly expensive.
We further investigate researchers' concerns with the application processes through our in-depth interviews.

\begin{table}[h!]
\begin{threeparttable}
\centering
\caption{Application outcomes by platform}
\begin{tabular}{lrrr}
\hline
\textbf{Platform} & \textbf{Applied} & \textbf{Denied} & \textbf{Accepted} \\
\hline
Bing & 3 & 0 & 0 \\
CrowdTangle & 21 & 2\tnote{a} & 9 \\
Google (Search \& Map) & 6 & 0 & 1\tnote{b} \\
LinkedIn & 2 & 0 & 0 \\
Meta Content Library & 19 & 1 & 6 \\
Snap & 2 & 0 & 0 \\
TikTok & 21 & 10 & 0 \\
Twitter & 31 & 9 & 9 \\
YouTube & 11 & 0 & 5 \\
\hline
\end{tabular}
\label{tab:applications}
\begin{tablenotes}
    \item[a] CrowdTangle was closed.
    \item[b] Only access to Google Maps was granted.
\end{tablenotes}
\end{threeparttable}
\end{table}

Next, we inquired about respondents' experiences with platform applications and their outcomes and list the results in Table~\ref{tab:applications}.
For those who did apply, we gathered detailed information about both the application processes and their outcomes.
Our findings reveal that the majority of applications were still pending at the time respondents completed our survey.
While many researchers had been waiting for at least a month since their initial application, some reported even longer periods of uncertainty.
Among those who received responses, a significant number faced rejection of their data access requests.
This trend was especially pronounced for platforms like X (Twitter) and TikTok.
Notably, when platforms rejected applications, they frequently did so without providing any explanation or justification for their decision.

\subsection{The Process of Permitted Access}

\begin{figure}[t]
    \centering
    \includegraphics[width=\linewidth]{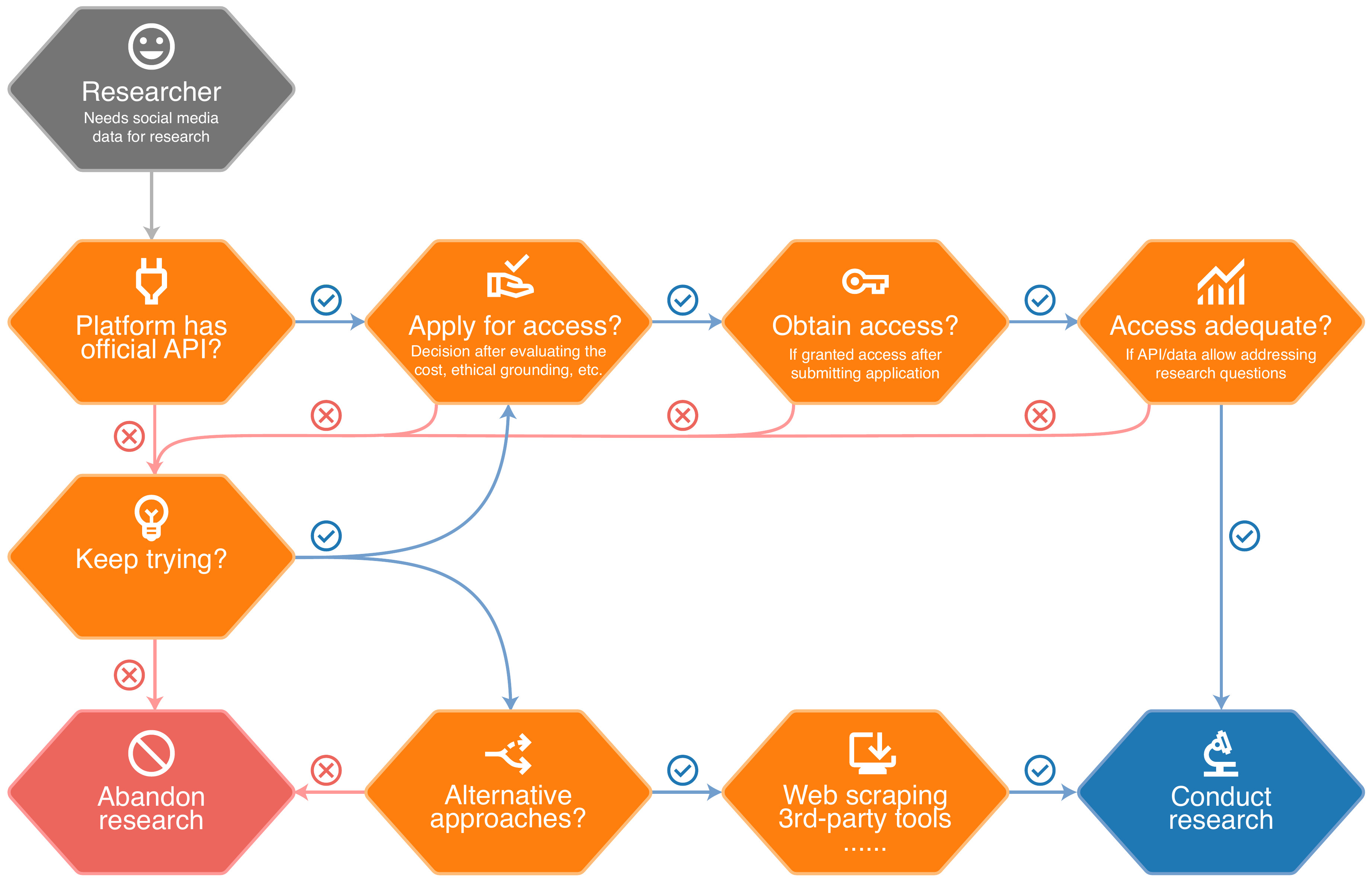}
    \caption{
      A flowchart demonstrating the steps and challenges researchers face when attempting to access social media data in the post-post-API era.
      }
    \label{fig:flowchart}
    \Description[A flowchart demonstrating the steps and challenges researchers face when attempting to access social media data in the post-post-API era.]{A flowchart demonstrating the steps and challenges researchers face when attempting to access social media data in the post-post-API era.}
\end{figure}

Our survey highlights several key barriers to data access for researchers.
To elaborate on these barriers and understand their impact on research, we rely on our interview data to contextualize these quantitative findings.
Using these combined results, we present a flowchart in Figure~\ref{fig:flowchart}, which highlights the many factors that may hinder researchers' ability to gain access to a platform's data program and ultimately cause them to abandon their research.
The flowchart illustrates four critical barriers researchers face: (1) lack of an official API or insufficient awareness of existing API access programs, (2) overly complex application processes, (3) significant delays by platforms or their third-party proxies in granting access, and (4) inadequate or limited data quality after access is provided.
To successfully conduct their desired study, researchers must navigate and overcome all these barriers. Otherwise, they will either have to resort to alternative data access approaches or abandon their research.

Our qualitative findings from our interviews highlight the nuanced and multifaceted nature of data access procedures, the challenges with widespread variation in application procedures, and the subsequent impact on researchers' willingness to apply for data at all.

\subsubsection{Barriers to Applying}
One of the barriers mentioned by researchers is the application process. Aligning with the survey results, some interviewees said they chose not to apply for API access for certain platforms because they found the application problematic.
This sentiment was most glaringly obvious when they spoke about X (Twitter).
Some researchers felt discouraged to apply or pay for the X API as they perceived the platform's business practices under Elon Musk as going against democratic values.
Bob, an information scientist in the U.S., decided to move away from research on X because he finds it ethically wrong to support the platform's business in any way.
\inlinequote{It's now hostile to my values, to research, to knowledge ...... it's no longer a great research site for me to study the things I want to study,} said Bob.

When researchers applied for APIs, they described the application as cumbersome, especially as each platform had a different API application process, and the experiences of the researchers varied.
Researchers lamented that the processes often did not reflect the realities of how research is conducted, were unnecessarily laborious, and required excessive time and resources.

For example, researchers are expected to apply for API access for individual research projects rather than for long-term access under the current data access regime.
Researchers felt that this application was not designed to facilitate flexible uses of APIs.
Max, a U.S.-based academic researcher who manages a large-scale research lab, explained that this requirement makes it hard for researchers to use the APIs simultaneously to conduct multiple projects.
Discussing Meta and TikTok, Max said:

\interviewquote{%
[The application processes were] kind of painful because they are adopting a model where they think that each researcher will apply for access for a specific research question.
That is very, very cumbersome for a lab like us where we have many researchers working at many research questions at any given time.
So it's very limiting.%
}{Max}

These challenges highlight a mismatch between current data access procedures and the pragmatic practice of conducting research using social media data: data access limited to individual projects makes it difficult for researchers to deftly study rapid and recent trends in the social media ecosystem.

Other barriers to access involved the arduous application process. For example, researchers who applied for the Meta Content Library API in the early phase said they were required to have ethical approval from Institutional Review Boards (IRB) for their projects. Although this requirement was later removed, the researchers indicated that it discouraged many scholars from applying because (1) such a practice hinders researcher independence and (2) data collection, preliminary data analysis, and the process of designing research often go hand in hand. Brian, a U.S.-based information scientist, welcomed Meta's decision to drop the IRB requirement, but added:
\inlinequote{[T]he fact that they even required it in the first place was a very limiting thing.}
Furthermore, for some social scientists who are not based in the U.S., the IRB requirement was simply difficult to meet because obtaining ethical approval is not a common practice in their countries.
Elaborating on this point, Fabio, a European political communication researcher, explained:
\inlinequote{In my university, like in many, many other universities in Europe, this kind of ethical committee [is] ...... not really common in social science.}

Other researchers mentioned that they struggled with a long list of questions and what they considered unrealistic criteria they were required to meet.
Bastien, a researcher at a European civil-society organization, who applied for the LinkedIn API, recalled being asked around 40 questions at the time of his application.
The platform \inlinequote{sets a very high bar} with its requirements, including what he perceived as overly strict privacy protection requirements for public data, said Bastien.
While he agreed that platforms needed to ask some of these questions to ensure ethical data usage, Bastien felt that \inlinequote{there are almost no organizations that would actually be able to meet all these criteria.}
He added, \inlinequote{the requirements around safety, security, privacy are absolutely disproportionate to the sensitivity of the [public] data.}

Crafting answers to these lengthy lists of questions can be highly time-consuming, which in turn slows down research.
Andy, a E.U.-based media studies scholar, described the application forms for Meta's tools as \inlinequote{pretty clear and straightforward} but felt that the applications were \inlinequote{unnecessarily onerous,} which \inlinequote{slows down your research project.}
Exacerbating the issue, aligning with the survey results, interviewees reported that it took them weeks, or sometimes months, before they heard back from platforms for their decisions.
While researchers who obtained TikTok API access reported what they considered a relatively short time frame before being able to use the API, that was not the case with Meta: they had to sign a legal agreement with the platform, which they described as a cumbersome process requiring legal assistance and considerable time resources. As Max put it:

\interviewquote{%
It's a little frustrating because, of course, there is a [legal] agreement that needs to be signed ...... every time there is an agreement, it has to go to the legal office of the university, and then they interact with the legal office at Meta.
Sometimes these interactions are quick and easy, and other times they stretch on for weeks or months ...... For the Meta Content Library, it probably took several weeks, maybe a few months.%
}{Max}

In sum, although API application processes and experiences vary by platforms and by researchers, our interviews highlight that the applications generally seem to have become much more time-consuming and complex under the current regime, requiring researchers to invest more resources.
This change made many researchers feel that platforms fell short of adequately supporting research endeavors.
For researchers not based in the U.S., some requirements, such as IRB, made the application process even more cumbersome, as these requirements seem not to reflect research practices at non-U.S. institutions.
As Ista, a E.U.-based computer scientist, put it: \inlinequote{They don't understand the needs of the research community well enough, or what they need to do to meet them.}

\subsubsection{Barriers to Access}
Unfortunately, many interviewees were denied access to research APIs after going through these application processes.
They either found the reasons for access denial provided by platforms unreasonable or felt that they were not given adequate explanations, believing that their research initiatives fell under what the European Commission calls research on systemic risks affecting E.U. citizens.
In worse cases, researchers did not receive any communication from the platforms on their application statuses.

The majority of the rejected cases in our interviews involved trying to access X data.
Erkki, a digital social scientist and a Ph.D. student based in the E.U., shared that his application for X's academic research access was declined.
Erkki has extensively studied climate discourses on X since before Elon Musk acquired the platform and applied for the X API to conduct research on how parliamentarians and municipal electoral candidates communicate about climate change.
Although he thought there would be no reason for his application to be declined because he had access to the previous versions of the Twitter API and used them for similar research projects, the platform denied him access on the grounds that his application was ``incomplete or lacks sufficient detail to show'' that he met the criteria under the DSA, according to the email from X Developer Team shared by Erkki.
Yet, \inlinequote{I don't really see any reason why they would reject it this time,} said Erkki, adding, \inlinequote{obviously it felt like an injustice has been done.}

Ista (AR/E.U.) also applied for X's academic research access with an eye to studying hate speech and counter-speech on the platform.
After exchanging some emails with the platform to provide requested additional information, her application was denied without a \inlinequote{very consistent reason}:

\interviewquote{%
[I]n the end, they denied [the application] with the reasoning that the research question doesn't fit the scope of the DSA, which I disagree with. But there is no process to actually contest the denial.%
}{Ista}

Some researchers never heard back from the platforms, which they  interpreted as effectivly a rejection.
The lack of transparency and communication from the platforms left researchers in limbo, without knowing whether they would be able to carry out their research projects.
Kay, a communication scholar at a public university in the U.S., applied for the Reddit API when it was first announced in 2023.
But she \inlinequote{never heard anything back,} said Kay, adding that such a case is \inlinequote{not unique to me.}
Corroborating Kay, Thirteen, a computer scientist in the U.S. studying social movements, said he only found out that his Reddit API application was rejected when he attended a workshop where a Reddit employee indicated that the decisions had been made.
The information was only \inlinequote{disclosed informally at the [workshop], not in a transparent, public forum,} said Thirteen, adding that he was unsure why his application was rejected.

Researchers who are not affiliated with universities face even more limited data access because some platforms prioritize academic researchers.
Different treatments of academic and non-academic researchers are not in alignment with Article 40 of DSA, which stipulates that researchers, including those at non-academic institutions, who are approved by national public authorities, must be granted access to platform data~\cite{techpolicyResearcherAccess}.

TikTok is one of the platforms that, at the time this research was conducted, excluded non-academic researchers from API access.
Bastien (NR/E.U.) applied for the TikTok API while appealing to the regulators of the country he was based in to obtain support for pushing TikTok to approve his application.
However, the application was declined on the grounds that he was not affiliated with a university:
\inlinequote{They told us, `right now, it's not available to non-academics, but we are planning on releasing it wider soon and we'll let you know.' They never let us know,} said Bastien.
He sent a follow-up email to TikTok, but received no response.
Similarly, Devin, a computational social scientist at a publicly-funded European research institution, was denied API access on the same grounds.
Devin thinks the reason for rejection is not justifiable because the Digital Service coordinators he had a meeting with made it clear that publicly-funded research institutions should be able to access data through platform APIs.
\inlinequote{We're not a university, it is true, but that's not a reason that we do not fall under the DSA,} said Devin.
This treatment of researchers not affiliated with universities is, in Bastien's words, a \inlinequote{mismatch between what they offer and what the law says they have to offer [under] the DSA.}

Another interviewee, Green Wave, a researcher at a civil society organization in Latin America providing support for marginalized populations targeted by harmful social media activities, also voiced frustration about TikTok's exclusion of non-academic researchers.
To save time and resources, she decided not to apply for the TikTok API, given that it was designed mainly for academic researchers in the U.S. and E.U.
However, not having access to this infrastructure affected her work, and she felt that platforms giving access to a limited pool of researchers went against the spirit of Article 40 DSA:

\interviewquote{%
The DSA was created to spot systemic risks that platforms created for society.
Civil society is playing an important and critical role in this because they ...... can see how different vulnerable communities are interacting with social platforms and service providers.
I really think that it's easier to give data access to the academics because they have the whole data process standardized.
But in terms of impact and making visible the systemic risks, civil society should have a role in it and a voice.%
}{GreenWave}

Stepping back, our interviews highlight that the decision-making processes regarding API applications are far from transparent.
In many rejected cases from X, the platform's justification was that these research projects fell outside the scope of the DSA, without elaborating on why or how.
Some researchers never received responses from the platforms, leaving them to speculate about the reasons for denial.
For researchers not affiliated with a university, access is even more limited because the door is closed from the beginning.
Although some researchers said they would try to apply again or find alternative ways to collect data, not having official API access can limit the amount and the kinds of data scholars can use, narrowing the scope of their research significantly (we will come back to this point later).

\subsubsection{Barriers to Usability}

Despite the challenges mentioned above, there are cases where researchers were granted API access.
When researchers gain access to APIs, what are their experiences? Are they satisfied with how the APIs function and the data they receive? Unfortunately, the answer is currently no---most of our interviewees reported encountering various obstacles while using the APIs.
We observed three primary issues that made it difficult for scholars to conduct research using APIs: (1) difficulty collecting data in the first place, (2) restrictive data caps, and (3) poor data quality.

\paragraph{Difficulty collecting data}
One major challenge researchers highlighted was the complexity and poor usability of APIs, which creates a significant hurdle to collecting data in the first place.
The platform most frequently mentioned was TikTok, aligning with previous reports~\cite{brown2023problem}.
Andy (AR/E.U.) found data collection through the TikTok API particularly challenging due to frequent server errors.
He explained that most of his data collection attempts have failed due to these errors, and he was \inlinequote{not able to get anything out of it.}
Other researchers provided detailed descriptions of these errors.
Fabio (AR/E.U.), who has been documenting the errors he has encountered while using the TikTok API, explained that he experienced multiple server errors, including ones that indicated the server received too many requests and problems with pagination.
\inlinequote{In theory, you should be able to download 100,000 videos a day, but in practice, I never even remotely reached that number,} said Fabio.
Max (AR/U.S.) corroborated these issues, noting that he also experienced similar server errors when using the TikTok API.
\inlinequote{[I]t's really not a very good quality product ...... a lot of the queries get errors in return. We have to try them several times} to collect data, said Max.

Researchers also described challenges in collecting data from other platforms.
In discussing the Meta Content Library API, Fabio (AR/E.U.) said the tool did not function as described in its public documentation, making data collection difficult.
He described the tool as \inlinequote{barely usable,} arguing that its design is \inlinequote{way too much focused on preserving any kind of risk of leaking the data} rather than facilitating research.
Elaborating on this point, he explained that users must navigate complicated, time-consuming steps before being able to use the API, including downloading a VPN software, enabling dual-factor authentication multiple times, and logging into virtual environments, to name a few.
Fabio said \inlinequote{I don't think anyone is actually using it ...... I think we're giving up with this third-party environment because it's not really usable.}

\paragraph{Restrictive data caps}
Even when researchers were able to collect some data, various data caps prevented them from obtaining adequate data to answer their research questions.
Researchers acknowledged that certain limitations are necessary, but found current data caps excessively restrictive.
One such limitation is the amount of data that can be retrieved.
In discussing Meta Content Library, Brian (AR/U.S.) said he has not \inlinequote{really used it} because of its restrictions: \inlinequote{It has very strong restrictions on the amount of data you can get at any given time ...... that make it virtually useless for the sort of work that I want to do,} explained Brian.
Although the tool allows users to pull 100,000 records at a time, he noted that \inlinequote{From a big data perspective, that's not a lot at all.}

Similarly, Thirteen (AR/U.S.) found the TikTok API's rate limit \inlinequote{not very generous.}
He explained that although he was trying to collect data from a relatively small number of TikTok accounts (about ``a few thousand''), what he deemed as strict rate limits made it \inlinequote{challenging to manage all the necessary data requests within the required refresh window.}
For large research teams, these restrictions mean multiple projects cannot run at the same time as they need to work under a single quota limit.
At Max's (AR/U.S.) research lab, \inlinequote{there's a single quota limit for all the people in the group} and \inlinequote{we have to make an agreement so that only one particular project can use the API at any given time because the quota is very, very limited,} said Max.

Another frequently-mentioned limitation was restricted access to certain data fields.
For instance, researchers, particularly those based in the E.U., reported that Meta's data access rules prevented them from retrieving information from profiles or pages with fewer than a certain number of followers.
This, according to researchers, posed a significant challenge for researchers studying countries with smaller populations than the United States, where Meta is headquartered.
Meta allows researchers to download Facebook content from pages with 15,000 or more likes/followers and profiles with 25,000 or more followers~\cite{facebookContentLibrary}. However, Bastien (NR/E.U.) said these thresholds are a \inlinequote{very high bar} for a country with a relatively small population he is interested in observing.
Frank, an E.U.-based media and communication researcher, also expressed frustration that these restrictions seem to be set based on U.S. standards, making them impractical for his research: \inlinequote{What's considered a major page or a big page on Facebook, or a big account on Instagram is something very different in the U.S.}
Brian(AR/U.S.) saw various restrictions on accessible data fields placed by platforms as a major obstacle in conducting novel research:

\interviewquote{%
They [platforms] are not doing the research.
They're always going to be designing the tool for last year's research ...... If you're trying to replicate what other researchers have done, you're not doing something novel anymore ...... [some data fields are] not there because somebody decided that that wasn't important and this other thing was important.
There's certain affordances of the tool that make it well suited to answer some questions and not others.%
}{Brian}

\paragraph{Poor data quality}
Interviewees also reported inaccuracy in API data, particularly with the TikTok API.
In alignment with the previous literature documenting notable discrepancies between API data and the TikTok website before July 2024~\cite{pearson2025beyond}, nearly all of the interviewees who have used the TikTok API reported either missing components or inconsistencies in terms of publicly available videos.
For example, Emily, a computational social scientist based in Europe, said the API often returned fewer, or sometimes more, transcripts of TikTok videos than were actually available on the TikTok app for a given time period.
She noticed this inconsistency when she compared the API data with datasets she purchased from a third-party vendor.
Emily added that the TikTok API is \inlinequote{not ideal} for research as she is required to do extra work to check inconsistencies and clean the API data.

Like Emily, other researchers shared that they have regularly experienced such inconsistencies with the TikTok API.
Frank (AR/E.U.) also manually compared posts of a politician collected from the TikTok API with what's displayed on the app, and found inaccurate results:

\interviewquote{%
I'll set up [the query to collect] all of [the politician's] TikToks from September of this year.
Run it, and it collects 10 TikToks.
And then I look at his profile, scroll, and try to manually check that it's correct.
He has 12 during September of that year.
I ...... run the API query again, and then it only gets nine.
So there's something weird happening with the API and it doesn't give consistent results.%
}{Frank}

Although the timeframes within which researchers collected their data vary and the platform seems to have been working on fixing these problems~\cite{pearson2025beyond}, interviewees found the API data highly unreliable and insufficient for conducting research.
As a result of these discrepancies, researchers ended up \inlinequote{not getting into it [using the API]} (Ista), having to find \inlinequote{another way to get that data} (Frank), or even \inlinequote{deprioritized my TikTok research} (Thirteen).
Thirteen (AR/U.S.) moved away from conducting research on TikTok because he and his colleagues \inlinequote{ultimately did not believe that we would be able to answer the research question} using TikTok API data.
The current state of data access made researchers feel that the platforms are simply trying to \inlinequote{check the box} (Frank) to meet the minimum regulatory requirements rather than genuinely supporting researchers. Many researchers expressed distrust in platform efforts, as Frank put it:

\interviewquote{%
I have a sneaking suspicion that platform owners, not only TikTok but X and Meta and all these different actors, they are obliged to provide something to researchers but what that something is and how it works or doesn't work isn't necessarily regulated. It just says they have to do something. So sometimes, my mind drifts into thinking that this is the bare minimum of what they can provide.%
}{Frank}

So far, we have described researchers' experiences of applying for APIs, cases of denied API applications, and the challenges they face in collecting adequate data through APIs.
Our interviews indicate that the application process and data access are in a constant state of flux, creating ambiguity.
While it takes time and resources for platforms to build reliable and sustainable research infrastructures, interviewees' experiences highlight that the current data regime has fallen significantly short of allowing researchers to conduct meaningful research under the DSA.

\subsection{Alternative Approaches}
\label{sec:alternative-approaches}
Given the many factors that limit researchers' ability to successfully traverse the process of permitted access, from submitting the application and gaining data access to using the data for research, our participants highlighted the need for individual and collective alternative data access approaches.

\subsubsection{Web Scraping}
Many participants noted that, despite changing data access procedures, there were still unofficial methods for collecting social media data. Peter, a European academic researcher, said, \inlinequote{if there are APIs, it's nice, else I will try to find my own way of getting data.}
Similarly, Sarah, a European digital studies researcher, noted that data collection is possible without APIs: \inlinequote{Python scripts where you can get Instagram data with. There's also still the possibility of in-browser scraping, which is done for Twitter.}
However, these alternative collection approaches also had limitations, such as \inlinequote{temporal gaps,} meaning \inlinequote{you cannot go into historical data well,} added Sarah.

One common alternative approach mentioned was data scraping.
Some platforms, such as Alphabet (which owns Google and YouTube), allow permissive scraping, as Bastien (NR/E.U.) was able to get permission for: \inlinequote{What they grant you is the right to scrape YouTube basically from a given IP address.}
However, in most cases, the legality and ethical ambiguity of scraping have raised concerns.
For example, Kate, a European computer science researcher, said, \inlinequote{There are a lot of scraping tools even now with which you could scrape Twitter, but I'm not sure if it's ethically okay to do it for research.}
In other words, even if researchers could scrape data (and in many cases, they were able to), researchers wanted to ensure that they did so legally and ethically.
Kay (AR/U.S.) said, \inlinequote{I will go to my grave saying scraping is not a crime, but it does go against the terms,} highlighting this difficult balance between doing important research on public data and adhering to terms of service obligations.
She said she still collected data, but set herself some boundaries: \inlinequote{I do still scrape web forums that are open, that don't require logins and websites.}
Researchers like Devin (NR/E.U.) also noted that they were more comfortable scraping data about aggregate information (such as engagement), rather than individual-user data: \inlinequote{We didn't look at user-level data. But instance-level data. So we scrape data. And that's also data that I think is safe to consider [a] public right.}

Others said they would avoid using scraping as a data access strategy altogether because it is not the most reliable means of collecting data.
For instance, Erkki (AR/E.U.) said he \inlinequote{really wouldn't trust} any longitudinal or historical research relying on scraping because it is hard to obtain comprehensive data.
Similarly, Fabio (AR/E.U.) said he is wary that the code used to scrape data could suddenly stop working because \inlinequote{when you are scraping, you are basically getting data in a way that is not supported by the platform.}
Fabio added that the issue is not only the lack of official platform support for scraping but also the barriers they could put up to stop scrapers (e.g., rate-limiting traffic, detecting and blocking bots, or even taking legal action).
Scraping means it is \inlinequote{not only not supported [by the platform] but sometimes actively fight [the platform],} said Fabio.

\subsubsection{Other Approaches}
There were many other alternatives that researchers considered other than scraping.
Some used third-party applications and tools, such as Zeeschuimer, a browser extension; Apify, a web-scraping and automation cloud platform; and Brandwatch, a social listening tool. Others used existing archives of past data.
Kate (AR/E.U.), for example, noted that she was using data she had collected previously: \inlinequote{I'm working with archival data that I have collected because it's no longer feasible for researchers to get data [through APIs].}
While these approaches have their own individual advantages and disadvantages, they collectively rely on some resources that are not democratically available to all researchers.
Barriers include money (to buy access or to pay participants), coding proficiency (to build or use a scraper), or the difficulty of obtaining comprehensive historical or longitudinal data.

Because alternative data collections are imperfect, but official access is not available, many expressed concern that important research was not being conducted.
Discussing the disadvantages and challenges of using alternative data access methods, Fabio (AR/E.U.) lamented, \inlinequote{You need to have a reliable way to assess this data ...... this is really unfortunate because we don't really have scraping [or other reliable and equitably accessible tools] and we don't really have the data from official channels.}
Bastien also expressed frustration that third-party vendors \inlinequote{don't necessarily have the data we need and the format we need} and such tools usually provide data that is collected in a \inlinequote{less systematic way than what we would have with the APIs.}

\subsection{Present Experience with the DSA}

Discussions about the DSA and its potential value are future-focused, with the acknowledgment that government regulation is only considered because platforms are not themselves willing to provide the data necessary to conduct independent research.
In general, the participants expressed  ``cautious optimism'' about the DSA: the future may rely on regulatory bodies to hold platforms to account, but it is not clear how this will actually occur and how data access will be provided.
For example, Peter (AR/E.U.) expressed optimism that platforms do not necessarily determine what access is or is not permissible:

\interviewquote{%
One of the crucial parts of the DSA  ......  is that the decision of who gets the data access and what kind of data should be given access to isn't with the platforms.
In that framework, it is with what's called the digital service coordinator.
So, basically, a government agency in one of the member countries.
I think that is a model that works a lot better for academic research because, of course, platforms have an interest in controlling the type of data they get you but also minimizing the amount of data they give and so on.%
}{Peter}

In this quote, Peter puts a lot of responsibility on the Digital Service Coordinator (DSC) as the ``third-party'' between what platforms are willing to provide and what researchers claim to want as part of their research.\footnote{DSCs assess the eligibility of researchers under DSA Article 40.4 that concerns private data but not 40.12, the subject of the public data access programs, see (\url{https://algorithmic-transparency.ec.europa.eu/news/faqs-dsa-data-access-researchers-2023-12-13_en}).}

When speaking more tangibly, researchers expressed frustration.
Erkki (AR/E.U.), in particular, highlighted the difference between applying (which could be easy) and the common rejection researchers experience. For example, he applied for Twitter access and had earlier versions of API access. However, his Twitter application under the DSA was rejected:

\interviewquote{%
To me, initially it looked pretty straightforward. I just fill a form and I send it. But, of course, the wordings and the context of some of the questions are different for the U.S. and for [the country he's based in] or basically any European nation. E.U. countries tend to have not politically aligned national institutions and research centers and such.
But the tone of the whole form, to me, indicated that it might be a mistake for me to apply with my status at the [non-academic research institute] instead of [the university he is affiliated with], which I did not consider ...... at that point.%
}{Erkki}
This quote highlights the minutia that can affect the acceptance or rejection of an access application.

Green Wave (NR/L.A.) saw some optimism in data access programs mandated by the DSA, noting that Article 40 DSA \inlinequote{is an interesting opportunity for researchers outside the European Union to maybe have some new possibilities in terms of data access.}
However, she also argued that there is a \inlinequote{transnational group disparity} between the Global South and the Global North as well as academia and civil society.
As a result, the gap between the data ``haves'' and the data ``have not's'' can widen under the DSA, particularly if they continue to disregard research outside of the academy.
These experiences suggest a dichotomy between what participants hope will come out of the DSA and what they are currently experiencing.

\subsection{Desired Future}

\subsubsection{Inflection Point to do Better Research}
Many interviewees shared the view that social media research is at a critical juncture with the loss of free APIs.
While the current conjuncture is often described as the \inlinequote{API Apocalypse} (Sarah, AR/E.U.), researchers also discussed ways to conduct better research within the current limitations, critically reflecting upon how the older data access regime has shaped their research fields.
Some researchers argued that social media research has tended to concentrate on platforms with easier access to data, which they said has led to overlooking other important avenues for communication.
They agreed that Twitter (X) was a popular platform for research, partly because data was readily available. Peter (AR/E.U.)'s following quote illustrates this point well:

\interviewquote{%
I think there was also a tendency for researchers to just do their research on Twitter because it was convenient and you could get a lot of data there.
And then that led to this research landscape where there was a lot of research that kind of took Twitter as a proxy for the internet, which I think is a problematic notion. So in a way, the fact that now we as researchers also need to maybe consider, okay, what platform should we be looking at?%
}{Peter}

Max (AR/U.S.) also saw the potential of producing more research on other relevant platforms as \inlinequote{the bright side} of the ongoing discourses around data access under the DSA.
\inlinequote{Twitter was the platform where this data was most available ...... but as people are moving to new platforms, especially because these platforms are open, then we hope we can do that again and use the kind of data on other platforms,} added Max.

Others argued that researchers should consider the possibility of small data research instead of focusing only on large datasets.
For instance, Sarah (AR/E.U.) said that, with small data, researchers can make sure the data is reliable by manually checking if there are any discrepancies in data collected through APIs or scrapers.
Although the loss of free APIs has limited the scope of research, it \inlinequote{also points out what you still can do even with small data ...... you're much more inclined to zoom into certain niches,} added Sarah. As these interviewees' quotes indicate, some researchers are exploring new opportunities for social media research despite the lack of adequate data access under the DSA.

\subsubsection{Collective Action and Policy}
At the same time, however, researchers emphasized that platforms are responsible for ensuring equitable and adequate data access and that the research community needs to collectively pressure the platforms to step up their efforts to support research.
They argued that platforms currently place strict limitations on who can access what kinds of data, as we have described above, and that they have not sufficiently considered the needs of the research community.

Brian (AR/U.S.) said that the current data access discourses are primarily driven by \inlinequote{the language of privacy, data minimization, protecting human subjects}, the rhetoric that platforms could weaponize to limit data access as much as they can.
While acknowledging that \inlinequote{all of that is obviously an important consideration,} Brian indicated that a healthy balance between these ethical issues and ensuring data access to conduct important research is critical.
\inlinequote{The bureaucracy is not serving the science [and] we're not able to actually answer meaningful questions} under the current framework, added Brian.
Researchers also voiced concerns that the current data access framework may compromise the independence of research from platform influence, calling for a community-wide discussion on the issue.

The current API application process that requires researchers to provide the details of their projects and the lack of transparency in platforms' decision-making processes were of particular concern among researchers, as platforms could easily deny API access to research that could put pressure on the platforms to change their practices.
Researchers should call on platforms to \inlinequote{make independent research possible within the regulatory framework,} said Sarah (AR/E.U.).
To that end, researchers said it is paramount to build a joint effort to lobby for more ethical and sustainable data access.
Fabio (AR/E.U.) argued that \inlinequote{the creation of a community of scholars and projects ...... is super important from this perspective, because I'm always dealing with platforms in a point-to-point relationship.}

\subsubsection{Types of API Data Desired}
As we have described in Section~\ref{sec:alternative-approaches}, researchers are eager to explore alternative data collection methods to overcome the current limitations of APIs, but there are significant limitations of such approaches.
For many social media scholars, reliable and sustainable API data access remains a desired means to conduct novel and creative research; researchers spoke about what types of data they wish they could access through APIs to answer their research questions.
In this subsection, we list the desired API data mentioned by our interviewees in the hope that it will serve as a step toward encouraging platform-researcher dialogue on improving data access.

One type of data often mentioned by researchers is information that is already publicly accessible or data on public figures (e.g., politicians, celebrities, and news organizations).
Researchers argued that platforms should be able to make public data readily available for all researchers, regardless of their affiliations, given that there is less concern about privacy.
Bastien (NR/E.U.) shared that he and his colleagues have mainly requested publicly available data from different platforms, but these attempts have been unsuccessful so far, adding that the bar is higher for non-academic researchers like him.
\inlinequote{Ideally, we would want to have access to an API that was broadly like the (older) Twitter API,} but if that's not possible, \inlinequote{what we would have liked was the possibility to input one account and get a list of all of their publicly accessible content for the last N years, along with a number of interactions,} added Bastien.
Emily (AR/E.U.), who studies TikTok posts of high-profile comedians, said platforms should loosen limitations on data related to public figures because \inlinequote{their data is in public domain} and it should be \inlinequote{OK for us to have it or even allow researchers to share the data with other scholars.} An important effort to develop protocol for determining what counts as publicly accessible content has been advanced by the Knight-Georgetown Institute's Gold Standard Working Group.\footnote{\url{https://kgi.georgetown.edu/expert-working-groups/gold-standard-expert-working-group/gold-standard-faq}}

Another request we heard frequently is access to larger-scale data.
Many researchers emphasized that large-scale data access is crucial for understanding trends, user behaviors, and platform activities (e.g., content moderation), and information flows on a platform-wide scale.
\inlinequote{Without APIs, I'm never going to be able to create very big data sets,} said Wellstone, a researcher at a for-profit company.
This limitation forces researchers to work with smaller samples that may not be representative, making it hard to draw reliable and generalizable conclusions regarding platforms.
For example, Sarah (AR/E.U.) said she has focused on small data research on TikTok and such an approach can contribute to important research findings, but scalability is hard to achieve without the API:

\interviewquote{%
The captions, the hashtags, the visuals, the sound is a very important part on TikTok where all these remixes and riffs and stuff is going on ...... now I can get at the data but not at scale. And this is where I would need the API for.%
}{Sarah}

Like Sarah, many other researchers discussed the types of API and data they would want from TikTok, in part because the platform has grown popular in many countries. This data included visual captions embedded in the videos (Emily, AR/E.U.), commenter identities that allow the creation of interaction networks based on a given content (Max, AR/U.S.), and all posts from a given account (Thirteen, AR/U.S.).

Overall, these findings highlight a dilemma: while researchers hope to rely less on platforms for data to secure independence, they also recognize limitations of alternative tools and the potential usefulness of official APIs.

\section{Discussion}

Our empirical findings have highlighted that, despite platforms' efforts to meet data transparency requirements mandated by the DSA, their practices vary greatly, and current data access programs are far from being adequate to facilitate research on digital platforms.
While scholars have previously performed API audits~\cite{pearson2025beyond}, our paper provides details of various challenges researchers face in different phases of navigating the permitted access across platforms, highlighting institutional, regional, and financial obstacles that exacerbate inequities in data access.

These results should be understood with some caveats in mind.
First, both our qualitative and quantitative studies relied on purposive sampling, and the experiences of the researchers involved in this study are not representative of any research field or region.
This also means our paper falls short of documenting the state of researcher data access on all VLOPs and VLOSEs equally.
However, our paper provides a broad overview of the current data access regime under the DSA, uncovering cross-platform trends and issues rather than a single platform, and encapsulates the experiences of researchers from different regions and backgrounds.

Second, we are still at the very beginning of the post-post-API era and the DSA regulatory framework has not been fully established yet.
This means that what we described in this paper are subject to constant change. However, these findings may help shape global regulations in the post-post-API age. This can contribute to the CSCW research community---which has relied on social media as an important data source---by documenting the transitional phases of the current data access regime and by providing a point of comparison for future research.

To encourage a future-oriented discussion and concrete actions toward improved data access for the CSCW community and beyond, we offer recommendations for platforms, researchers, and policy based on our participants' experiences documented in this paper.

\subsection{Recommendations for Platforms}

First and foremost, platforms should ensure transparency in the API application process.
While previous research has identified inaccuracy in API data that hinders research~\cite{pearson2025beyond, corso2024we}, our findings highlighted that obtaining API access itself remains a hurdle.
Many researchers have been denied API access without detailed justifications or adequate communication from the platforms.
Based on our interviews, many cases appear to fall under the DSA's scope according to both our analysis and participants' perspectives. However, the lack of clear guidelines and communication makes it difficult for researchers to properly frame their data access requests or determine whether reapplication would be worthwhile.
For instance, while the DSA stipulates that data access should be available for research on systemic risks in Europe, there remains significant uncertainty around eligibility criteria---both in terms of which researchers qualify and what types of research projects are considered valid~\cite{arxivGreatData}.
Furthermore, as our interviews surfaced, researchers are expected to request specific data for their projects, but they do not know what specific data is available~\cite{arxivGreatData}.

While these issues stem partly from the DSA's operational limitations, platforms' lack of coherent application procedures and ambiguous decision-making have led researchers to suspect intentional suppression of unfavorable research while only meeting minimal regulatory requirements.
This has further eroded researchers' already fragile trust in platform-provided data access~\cite{allen2021research}.
Given that platforms maintain complete control over data access permissions, increased transparency and accountability are essential to foster productive researcher-platform dialogue and establish ethical data access practices that genuinely serve the public interest.

We also recommend that platforms consider simplifying the application design to facilitate more flexible uses of APIs. For example, under the current data access programs, researchers can only apply for temporary, project-based access, which some participants said significantly limits their ability to complete research, implement longer-term projects, or share the API access within a research lab.
It is understandable that platforms are reluctant to provide unlimited access to their data to avoid overburden, but more flexible access frameworks should be possible (at least in theory) given that older programs such as Twitter API and Meta's CrowdTangle featured greater versatility.
Ideally, platforms should allow researchers to apply for more long-term data access, rather than project-based access, so that researchers can conduct multiple research projects simultaneously as long as the overall research endeavor falls under the scope of the DSA.

Additionally, we recommend that platforms consider a more realistic balance between privacy protection and research.
While the current design of API data access programs appears to be primarily driven by privacy considerations, participants saw the privacy-related criteria as disproportionately strict to the sensitivity of publicly accessible data.
This approach is, unfortunately, discouraging many researchers from applying and preventing wider data access. As participants argued, platforms should consider less strict limitations on access to public data in terms of both quota and available data points.

For ethical and practical data access frameworks, platforms should engage in direct conversations on these issues with researchers from different disciplines and regions.
An example of this is Twitter's now-disbanded Academic Research advisory board that served to facilitate dialogue between researchers and the platform~\citep{blakey2024day}.
In a more recent development, it is promising to see that Reddit has worked directly with a small group of researchers to gather their feedback to improve the Reddit for Researchers (R4R) Beta Program~\cite{u/PeerRevue}.
Although Reddit is not classified as a VLOP, such an initiative is a promising first step that can set a model example. We argue that VLOPs and VLOSEs should invest more resources into supporting research by listening to researchers' needs because, ultimately, doing so will contribute to improving the safety and functionality of their own platforms.

\subsection{Recommendations for Researchers}

Over the last decade, research in HCI and CSCW has increasingly turned to social media data as a primary source to explore online social dynamics and human behavior, contributing to the design of new technological tools and interventions~\citep{alvarado2025knowledge}.
To continue building on this significant body of research, it is paramount that researchers foster interdisciplinary coalitions to collectively advocate for better researcher access to platform data.
Our interviews highlighted the need for more organized efforts to improve data access despite individual researchers' attempts to appeal or work with regulators and platforms.
Fortunately, in addition to research that examined API performance, there are ongoing initiatives working toward informed policies and ethical, independent data access.
To name a few examples, the Knight-Georgetown Institute\footnote{\url{https://kgi.georgetown.edu}} focuses on developing consistent standards for accessing publicly available platform data under DSA Article 40, offering recommendations to regulators~\cite{chapman2024laying}.
Coalition for Independent Technology Research\footnote{\url{https://independenttechresearch.org}} has been actively building an interdisciplinary coalition involving not only academics but journalists, civil society organizations, and community scientists to advocate for data transparency and ethical research.

As underscored in this paper, there are currently notable disparities in data access between academic and non-academic researchers; E.U.-based researchers and those who are outside of the region, particularly in the Global South; and those with funding and those without.
Given that third-party tools and other alternatives like scraping are neither adequate nor affordable for everyone, the institutional, geopolitical, and financial data access gaps can further increase among these groups without sustainable and equal access to official API data, and researchers, particularly those affiliated with academic institutions, should aim for inclusive dialogue.

At the same time, researchers should remain flexible and continue exploring different data collection approaches rather than focusing solely on the past models of data access.
The data collection methods discussed thus far---APIs, scraping, and other alternatives---are fundamentally platform-centric, as they rely heavily on platforms' data access policies and infrastructure.
However, an alternative paradigm has been gaining momentum: user-sourced data collection, which encompasses both data donation and tracking approaches~\cite{ohme2024digital}.
In the data donation model, users exercise their right to data portability (guaranteed under the General Data Protection Regulation) to download their personal platform data and voluntarily share it with researchers~\cite{carriere2025best}.
This approach has been facilitated by major platforms implementing data export functionality to comply with regulations.
The tracking model represents another user-centric approach, typically involving browser extensions that monitor users' platform interactions and data usage patterns~\cite{christner2022automated}.
Projects like the National Internet Observatory are developing robust infrastructure to support such tracking-based data collection and sharing among researchers under enhanced ethical frameworks~\cite{feal2024introduction}.
While these user-sourced methods face their own set of challenges, they enable novel research directions, particularly in understanding user engagement and behavior patterns, that were previously impossible with traditional platform-centric approaches.

\subsection{Recommendations for Policymakers}
Building researcher coalitions and collectively lobbying for improved API data access is important, but clear regulatory frameworks are necessary when platforms refuse to respond swiftly and adequately.
This is where we stand in the post-post-API era. We argue that more regulation is an inevitable step forward, particularly for independent research that holds these platforms responsible.

In Europe, as part of the DSA, the European Commission announced the Delegated Regulation to provide more clarity on how this data transparency should work in practice, focusing on specifying ``the conditions under which sharing of data should take place and, the purposes for which the data may be used''~\cite{europaEuropeanCommission}. This is an important step to address ``critical legal interpretational and operational challenges''~\cite{arxivGreatData} that have undermined the realization of data transparency, the regulation aims for.

While this move towards more evidence-based policymaking potentially addresses the arbitrary nature and the lack of transparency in platforms' decisions on data access permission, as well as the lack of consistency across platform data-sharing practices, policies should also support plurality in data collection approaches.
Beyond complying with basic regulatory requirements, platforms have little motivation to drastically expand the sharing of their data with researchers when they can profit from selling this data or leverage it to power their own generative AI tools.

Bearing this in mind, we argue that regulatory bodies should encourage not only permission-based data access but also more independent data access procedures.
For example, it may be fruitful to consider the possibility of a regulatory framework that allows scraping of public data, especially when platforms fall short of providing necessary data.
As described in this paper, scraping is one of the alternative data collection methods some researchers have relied on, but litigation attempts, particularly by Elon Musk~\cite{Bond_2023}, have created chilling effects among researchers wary of violating platform terms of services.
Policies supporting alternative data collection methods can help research that is critical of platforms without the fear of suddenly losing data access or resources.

The issue of data access becomes even more complicated from a global perspective.
Social media platforms and search engines operate on a global scale and, ideally, their data should be accessible to researchers regardless of language or region.
In reality, however, there is a divergence in platforms' regional policies, with Europe seeing more promising developments with the introduction of the GDPR (General Data Protection Regulation) and the DSA, as well as regulatory bodies like Ofcom.
This means Europe may lead empirical social media research going forward, while other regions, particularly the Global South, may be left behind, exacerbating existing imbalances in access to resources.
Platforms' data access policies are ephemeral, and it is neither realistic nor desirable to allow platforms to set the standards for data access.
While it is unclear how much data access can be achieved under the GDPR and the DSA, these policies can be a potential model for other parts of the world.

\section{Conclusion}
In this study, we used a mixed-method approach to understand researchers' experiences with digital platform data access, platforms' willingness to provide data in the post-post API age, and the necessary data for studying platforms robustly and transparently.  
New programs and tools mandated by the DSA initially garnered some hope among research communities, but our study indicates that they are far from providing wide researcher data access, nor do they provide data with necessary quality or quantity for emperical research.
While it is encouraging to see some platforms try to improve API access, and regulatory efforts to increase clarity in how these data access programs are enacted, we are experiencing a significant setback from the age of mass data access.

Without mincing words, research about social media and digital platforms is in dire straits, particularly for scholarship that seeks to hold platform accountable. What researchers are emperically able to study now largely depends on regulatory frameworks, like the DSA. However, it is unclear how these policies will be enacted practically, or whether it will hinder important and alternative approaches to data collection, including scraping and user-sourced data collection. While these approaches are beyond the scope of what the DSA can consider, they are nevertheless important mechanisms for which researchers are able to gather publicly-accessible and/or user-provided data. To encourage platform and regulatory efforts based on open and direct engagement with the research community and user base, further research and case studies are needed to document transitional phases of the DSA's data access regime.


\bibliographystyle{ACM-Reference-Format}
\bibliography{ref}

\appendix

\section{Participant Information}

\begin{table}
\begin{threeparttable}
\caption{Participant information.} \label{tab:participant-info}
\begin{tabular}{llllp{5.5cm}}
\hline
\textbf{Pseudonym} & \textbf{Pronoun} & \textbf{Affiliation}\tnote{a} & \textbf{Region}\tnote{b} & \textbf{Research Field(s)} \\
\hline
Andy & he/him & AR & E.U. & Media Studies \\
Bob & he/him & AR & U.S. & Information Sciences, Linguistics, Political Science, Sociology \\
Brian & he/him & AR & U.S. & Information Sciences, Psychology \\
Emily & she/her & AR & E.U. & Humanities, Information Sciences, Linguistics, Political Science, Sociology \\
Erkki & he/him & AR & E.U. & Communication, Computer Science, Information Sciences \\
Fabio & he/him & AR & E.U. & Communication, Political Science \\
Frank & he/him & AR & E.U. & Media and Communication \\
Ista & she/her & AR & E.U. & Computer Science, Sociology \\
Kate & she/her & AR & E.U. & Computer Science, Information Sciences, Psychology \\
Kay & she/her & AR & U.S. & Communication \\
Max & he/him & AR & U.S. & Computer Science \\
Peter & he/him & AR & E.U. & New Media Studies \\
Sarah & she/her & AR & E.U. & Communication, Humanities, Sociology \\
Thirteen & he/him & AR & U.S. & Computer Science, Information Sciences \\
Bastien & he/him & NR\tnote{1} & E.U. & Information Sciences \\
Devin & he/him & NR\tnote{2} & E.U. & Communication, Psychology \\
Green Wave & she/her & NR\tnote{1} & L.A. & Political Science \\
Philipp & he/him & NR & E.U. & Computational Social Science \\
Wellstone & Dr. & NR\tnote{3}  & U.S. & Business, Communication, Computer Science, Humanities, Information Sciences, Linguistics, Political Science \\
\hline
\end{tabular}
\begin{tablenotes}
\item[a] AR = Academic Researcher; NR = Non-academic Researcher
\item[b] E.U. = European Union; U.S. = United States; L.A. = Latin America
\item[1] Civil-society organization
\item[2] Public-funded research institute
\item[3] For-profit company
\end{tablenotes}
\end{threeparttable}
\end{table}

In Table~\ref{tab:participant-info}, we provide detailed information about the participants in our study.

\end{document}